\documentclass[letterpaper, 10 pt, conference]{ieeeconf}
\IEEEoverridecommandlockouts
\usepackage{cite}
\usepackage{graphicx}
\usepackage{textcomp}
\usepackage{xcolor}
\def\BibTeX{{\rm B\kern-.05em{\sc i\kern-.025em b}\kern-.08em
    T\kern-.1667em\lower.7ex\hbox{E}\kern-.125emX}}

\usepackage{colortbl,hhline}
\usepackage{amsmath}
\usepackage{amsfonts}
\usepackage{amssymb}
\usepackage{amsthm}
\usepackage{mathrsfs}
\usepackage{xspace}
\usepackage{xparse}
\usepackage{mathtools}
\usepackage{makecell}

\usepackage{algorithm}
\usepackage{algpseudocode}

\usepackage{booktabs}

\usepackage{epstopdf}
\usepackage{lipsum}
\usepackage{color, soul}

\usepackage{hyperref}

\usepackage{cleveref}

\usepackage[english]{babel}

\usepackage{booktabs} 

\title{Characterizing Human Feedback-Based Control\\in Naturalistic Driving Interactions\\via Gaussian Process Regression with Linear Feedback}

\author{
Rachel~DiPirro,
Rosalyn Devonport, 
Dan Calderone,  
Chishang~``Mario''~Yang,
Wendy Ju,
Meeko Oishi
\thanks{
This material is based upon work supported by by the National Science Foundation under NSF Grants No. 2227338, IIS-2107111 (Cultural Differences in Driving Interaction), and IIS-2212431 (Cultural Differences in Pedestrian-AV Interaction).
Any opinions, findings, and conclusions or recommendations expressed in this material are those of the authors and do not necessarily reflect the views of the National Science Foundation.
This work is also supported by the Jacobs Technion-Cornell global collaboration funds.
}
\thanks{Rachel DiPirro, Rosalyn Devonport, Dan Calderone, and Meeko Oishi are with Electrical
  \& Comp. Eng., Univ. of New Mexico, Abq., NM. Chishang ``Mario'' Yang and Wendy Ju are with Information Science, Cornell Tech, New York, NY 10044.
  Email: {\tt\{rdipirro,devonport,dcalderone,oishi\}@unm.edu}, {\tt\{cy546,wendyju\}@cornell.edu}.
}
}

\begin{document}

\maketitle

\begin{abstract}
Understanding driver interactions is critical to designing autonomous vehicles to interoperate safely with human-driven cars.
We consider the impact of these interactions on the policies drivers employ when navigating unsigned intersections in a driving simulator.
The simulator allows the collection of naturalistic decision-making and behavior data in a controlled environment.
Using these data, we model the human driver responses as state-based feedback controllers learned via Gaussian Process regression methods.
We compute the feedback gain of the controller using a weighted combination of linear and nonlinear priors. 
We then analyze how the individual gains are reflected in driver behavior.
We also assess differences in these controllers across populations of drivers. 
Our work in data-driven analyses of how drivers determine their policies can facilitate future work in the design of socially responsive autonomy for vehicles.
\end{abstract}

\section{Introduction}

A key element of designing effective autonomous driving behavior is the ability to capture interactions between vehicles.  Interactions capture nuances of inference, prediction, and action that are key for safe and effective autonomy.  
Numerous incidents and accidents have occurred in scenarios in which an autonomous vehicle (AV) violates expectations of the human driver, by enacting actions outside of what humans drivers would typically do.  The design of effective vehicle autonomy requires models that can capture and ideally mimic human driver actions, facilitating interaction by reducing opportunities for misinterpretation and confusion, and enabling interpretability.  Such capabilities can promote socially responsive AVs.

We seek to develop models of driver feedback laws during human-human driving interactions, that have structure which we can interpret, analyze, and tune as needed.  We focus on structured models, in contrast to black-box models such as neural nets, because they can function in scenarios in which data has not been previously observed, and because they let us discern and interpret potentially subtle differences that may have outsized importance for human drivers.  The type of model is non-trivial -- not only because of ease of analysis, computation, and interpretation, but also because presuming a type of model that is incompatible with the underlying model can result in further inaccuracies and interfere fundamentally with the feasibility of the model.  

We seek a modeling approach for human driving feedback laws for interactions that can 1) determine whether a linear model is adequate, 2) infer that model from previously observed data, and 3) be interpreted in terms of analysis of gains, just as a typical feedback law could.  
There is precedent for linear models of human driving feedback laws, such as the longstanding crossover model \cite{Mcruer}, which has been used to model manual control in cars and in aircraft as a delayed integrator, but just  in the context of specific, narrowly defined tasks (i.e., regulating altitude, manual driver overtaking and passing maneuvers) with a single vehicle.
More recent work has inferred and predicted the likely actions of drivers \cite{sadigh2014data, narang2023multiplayer, fox2018should, amsalu2016driver} via inverse reinforcement learning for the purpose of making autonomous vehicles responsive to human drivers.  
Other work has focused on interactions between a human driver and an autonomous agent
\cite{sadigh2016planning, sadigh2018planning, zhao2023measuring}, and on large-scale naturalistic data \cite{kuefler2017imitating,driggs-campbell2017integrating}, however, these approaches may not capture the subtleties of interactions between human-driven vehicles \cite{markkula2020defining}.
Data-driven and statistical models have also been used to capture policies from human behaviors \cite{6774467, 9258484, sadigh2017active, 6111421}, but these lack the predictability and interpretability that mathematical models of feedback laws provide.

The main contribution of this paper is the \textit{creation and validation of a linear feedback model of human driving policies in naturalistic driving interactions}.  
We propose a statistical methodology that not only allows for inference of linear behaviors, but that is also capable of determining whether a linear model is sufficient to explain the observations.  Specifically, our approach is based in Gaussian Process (GP) regression \cite{Rasmussen_Williams_2008}, which uses a combination of linear and nonlinear priors in a weighted sum where the weighting factors are decided by maximum likelihood.
Our approach is premised on the idea that we can compute a gain matrix from the linear component of the kernel.
This approach has the advantage that the resulting control gain is interpretable, in that the gains can indicate which parts of the state space are of high impact to the policy, and can be analyzed across populations.

This paper is organized as follows. 
We describe the experimental setup used to collect the naturalistic interaction data in section \ref{experiment}.
Section \ref{methods} describes the statistical model, the state-feedback controller, and tools employed for controller analysis.
Section \ref{results} presents our results and interprets our findings, and we offer conclusions in Section \ref{conclusion}. 

\section{Experimental Setup}
\label{experiment}

\begin{figure}[t]
    \centering
    \includegraphics[trim=14.8cm 0 14.8cm 0, clip, width=0.45\columnwidth, keepaspectratio]{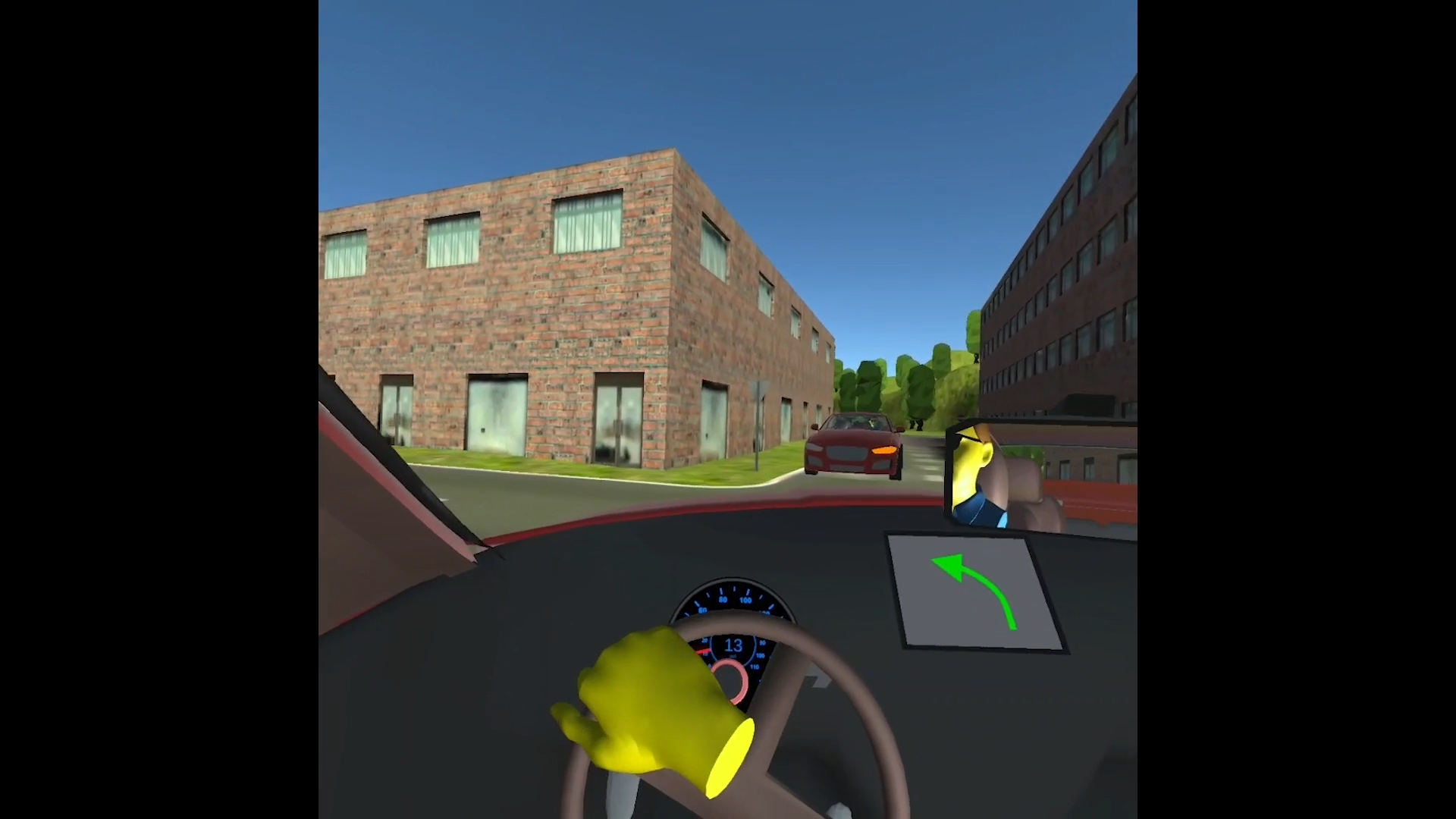}
    \includegraphics[width=0.45\linewidth]{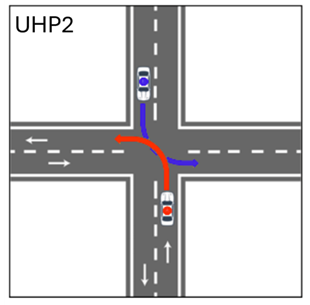}
    \caption{
    Left: Within the virtual environment, each driver has a realistic driving view that encompasses the dashboard of their own virtual car, the simulated driving environment, and the other vehicle.
    The green arrow on the driver's dash indicates the maneuver to be taken through the intersection.
    Right: We consider a single intersection scenario in this analysis, in which both cars are instructed to turn left. Car A is denoted in red and Car B is denoted in blue.}
    \label{fig:scenario}
\end{figure}

For this analysis, we consider a dataset focused on capturing interactions between human drivers.
These data were gathered using the StrangeLand driving simulator \cite{Goedicke_Zolkov_Friedman_Wise_Parush_Ju_2022}, which specifically enables multi-participant driving scenarios.
This simulator enables two drivers wearing VR headsets to control their respective simulated vehicles in a shared virtual driving environment using a physical steering wheel interface, as well as gas and brake pedal controllers.  
Each driver is able to see and respond to the other driver's actions within the shared virtual environment, as shown in Fig. \ref{fig:scenario}.

The scenarios are set up to facilitate naturalistic interactions between the drivers.
We consider an ambiguous right-of-way scenario involving an unsigned four-way intersection.
Each driver is given a specified maneuver to perform at the intersection by a navigation interface on the virtual vehicle's center console; however, the driver is able to drive freely within the simulator.
Timed traffic lights before the intersection are used as traffic control to improve the likelihood of interaction at the intersection \cite{schindler_dynamic_2016}.

The study was conducted in multiple locations.
A total of 170 participants were included, comprising 85 dyads: 42 in Israel (ISR) and 43 in New York (NYC).
Further information about the demographics of the participants can be found in \cite{10.1145/3640792.3675717}.
Each pair of participants completed seven intersection traffic scenarios with ambiguous right-of-way.
For this work's analyses, we consider only one of these scenarios, in which one driver approaches the intersection from the north, the other from the south, and both are instructed to complete left turns at the intersection, as shown in Fig. \ref{fig:scenario}.
This scenario is denoted as an unconstrained heads-on path, or ``UHP'', scenario as per the conventions in \cite{markkula2020defining}, and is referenced as ``UHP2'' in our prior work \cite{10919603}. 

The simulator records data at approximately 18 frames per second to ensure smooth performance in the simulator and includes information such as car location, velocity, acceleration, and heading.
Human experimental data were also collected at the same frame rate and include the angle of the steering wheel, gas and brake pedal information, hand position and orientation, and head position and orientation.
For our interests, we consider the vehicle data and the input data in order to model a linear feedback system.

\section{Methods}
\label{methods}
We propose a statistical model for functional regression that has three properties: (i) the contribution of linear and nonlinear parts to the overall goodness of fit must be clear, governed by interpretable parameters; (ii) there are as few \emph{a priori} restrictions on the nonlinear part as possible; and (iii) the degree of contribution of each part must be decided by a reasonable statistical procedure. Many popular regression models can be made to satisfy two of these, but few can satisfy all three. Generalized linear regression may satisfy (i) and (iii) but not (ii); a neural network model (augmented with a linear bias) may satisfy (ii) but not (i), nor (iii) to an extent; reproducing kernel Hilbert space interpolation satisfies (i) and (ii) but not (iii). 

Our approach employs a GP regression model with a prior covariance specially designed to balance linear and nonlinear terms. 
The relative dominance of the linear part is represented by a scalar hyperparameter, satisfying (i); the nonparametric nature of GP regression allows us to add a term to satisfy (ii); and the value of the weighting term in (i) may be selected according to an empirical Bayes procedure wherein the linear/nonlinear weighting factor is selected to maximize the marginal likelihood of the observations.

\subsection{Preliminaries}
Gaussian process regression models are a class of nonparametric Bayesian statistical function estimators in which prior information about the structure of the unknown function is encoded through the choice of a prior covariance function. Fixing a prior covariance function
$k:\mathcal{X}\times\mathcal{X}\to\mathbb{R}$ and a scalar observation noise variance $\sigma_n \ge 0$, a set of input data $x_1,\dotsc,x_N\in\mathcal{X}$, and a set of output data $y_1,\dotsc,y_N\in\mathbb{R}$, the GP regression model yields a distributional estimate of a function $f:\mathcal{X}\to\mathbb{R}$ hypothesized to relate the input data to the output observations as $y_i=f(x_i) + \epsilon_i$ where $\epsilon_i\sim\mathcal{N}(0,\sigma_n^2)$. To each point $x_*\in\mathcal{X}$ the model assigns a Gaussian random variable characterizing the likely values of $f(x_*)$; for our analysis we will only use the mean of these random variables, yielding a function estimate $\hat{f}:\mathcal{X}\to\mathbb{R}$ defined pointwise as 
\begin{equation}
\label{eq:gp_mean_generic}
\hat{f}(x_*) = \mathbb{E}[f(x_*)] = \sum_{i=1}^N \alpha_i k(x_i, x_*)
\end{equation}
where $\alpha = (K + \sigma_n^2 I)^{-1} y$ with $K\in\mathcal{R}^{N\times N}$ defined componentwise as $K_{ij} = k(x_i,x_j)$.
The matrix $K$, called the kernel Gramian, is symmetric and positive semidefinite, which ensures that $K+\sigma_n^2 I$ is nonsingular.

Since $\hat{f}$ is a linear combination of functions of the form $k(x_i,\cdot)$ it follows that the functional structure of $\hat{f}$ is determined entirely by the functional structure of $k$ with its first argument fixed. For instance, if $\mathcal{X}=\mathbb{R}^n$ and $k(x,y) = \langle x, y \rangle$ then $\hat{f}$ is a linear function, namely 
\begin{equation}
\label{eqn:f-hat}
    \hat{f}(x_*) = g^\top x_*
\end{equation}
with $g=\sum_{i=1}^N\alpha_i x_i$. Frequently we wish to place as few restrictions on the form of $\hat{f}$ as possible; for this purpose we may employ a so-called universal covariance function such as the squared exponential covariance $k(x,y) = \exp(-\tfrac{1}{2\ell^2}\| x - y \|^2)$. Unlike the inner-product covariance function, the squared exponential covariance contains a hyperparameter,
namely the lengthscale $\ell^2$.
In general there may be several hyperparameters,
which we collect into the vector $\Theta$ and denote the hyper-parametrized covariance function as $k_\Theta$; then the marginal likelihood of the observations for a given $\Theta$ is given by
\begin{equation}
    L(\Theta) = -\tfrac{1}{2} y^\top K_\Theta^{-1} y - \tfrac{1}{2} \log|K_\Theta| - \tfrac{N}{2}\log2\pi
\end{equation}
where $K_\theta\in\mathbb{R}^{N\times N}$ is defined componentwise as $(K_\Theta)_{ij} = k_\theta(x_i, x_j)$. In general $L$ will have multiple local maxima, meaning that the maximum likelihood optimization must be carried out with a nonlinear programming solver.

\subsection{Dynamical Model}

We consider the interactions between drivers as modeled by nonlinear dynamics with the form
\begin{align}
\label{eq:abstract_driver_dynamics}
\dot{x} = f(x,u_p), \quad x_{0} \ \ \text{given}
\end{align}
for each agent $p \in \{A,B\}$ corresponding to Car A and Car B respectively, the joint state space $x\in \mathbb{R}^{2n_p}$ and a single agent's controls $u_p \in\mathbb{R}^{m_p}$.
In the mathematical model, the proposed hypothesis of a linear feedback law is justified by a first-order approximation of the nonlinear interaction
\begin{align}
\dot{x}
& \approx f(\bar{x},\bar{u}_p) + \left[\frac{\partial f}{\partial x}\right]_{\bar{x},\bar{u}_p}\Delta x
+
\left[\frac{\partial f}{\partial u_p}\right]_{\bar{x},\bar{u}_p} \Delta u_p 
\label{eqn:dynamics}
\end{align}
to be valid in a region of $X\times U$ sufficiently large to encompass a reasonable fraction of the observations. Under these conditions a linear feedback law is a reasonable control strategy.
We do not, however, restrict our analysis \emph{a priori} to only linear models; indeed, our purpose in using a GP regression model is to allow the data to determine whether a linear model is appropriate.

\subsection{Gaussian Process Regression}
Since the goal of our statistical analysis is (i) to determine if the driver interactions are well modeled by linear terms and (ii) if so to recover the corresponding feedback gains, the prior covariance function for our GP regression model should include a combination of linear and nonlinear terms augmented by hyperparameters that scale the relative degree of their contribution to the overall data fit. A reasonable candidate for such a prior covariance is the function
\begin{equation}
k(x,y) =
   \theta_1 \langle x,y \rangle 
   + 
   \theta_2 e^{-\tfrac{1}{2\ell^2}\|x-y\|^2},
\end{equation}
which gives a weighted balance between linear regression from the inner-product term and universal GP approximation from the squared-exponential term. We can divide through by one of the parameters to reduce the burden of numerical maximum likelihood optimization; dividing through by $\theta_1$ and re-labeling $\beta = \theta_2/\theta_1$ yields
\begin{equation}
   k(x,y) = \langle x,y \rangle 
   + 
   \beta e^{-\tfrac{1}{2\ell^2}\|x-y\|^2},
\end{equation}
which is the prior covariance we use in our analysis.
The weighted sum of terms in the prior yields an estimator of a similar weighted-sum form, namely
\begin{equation}
    \label{eq:lin_se_regression}
    \hat{f}(x_*)
    =
    G^\top x
    +
    \beta \sum_{i=1}^N \alpha_i 
    e^{-\tfrac{1}{2\ell^2}\|x_i-x_*\|^2}
\end{equation}
as described in equations~\eqref{eq:gp_mean_generic} and~\eqref{eqn:f-hat}. 
This expression for $\hat{f}$ justifies our interpretation of $\beta$ as a \emph{nonlinear inclusion factor} that indicates the degree to which nonlinear terms are required to construct a reasonable explanation of the observations. In the extreme case of $\beta=0$ the estimator reduces to conventional linear regression; as $\beta$ increases, the terms arising from the universal GP approximator eventually begin to dominate. In our analysis, we use the maximum likelihood principle to justify selecting $\Theta=(\beta,\ell)$ with respect to a local maximum of $L(\Theta)$.

\subsection{State-Feedback Controller}
We formulate the linear state-based feedback controller for each agent $p$ in an interaction as 
\begin{equation}
    u_p = \bar{u}_p + G^Tz
    \label{eqn:feedback-law}
\end{equation}
where $\bar{u}_p$ is a nominal set of controls, $G$ is the gain matrix, and $z$ is a set of preselected state features.
This structure follows from the linearization in equation~\eqref{eqn:dynamics}, where we consider the approximation of the nonlinear interaction as nominal term plus a linear component.
The gain matrix $G$ in equation~\eqref{eqn:feedback-law}
is the same as the gain matrix in equation~\eqref{eq:lin_se_regression}, learned via GP regression.
We can presume that $G$ is a linear gain matrix, so long as $\beta < 1$, suggesting that more of the structure of the data is captured by the linear component of $G^Tx$ in equation~\eqref{eq:lin_se_regression}.
Future work will consider the impact of both $\beta > 1$, which implies that the data is mostly nonlinear, and whether a different threshold for $\beta$ could yield different results.

\begin{figure}[t]
    \centering
    \includegraphics[width=0.5\columnwidth, keepaspectratio]{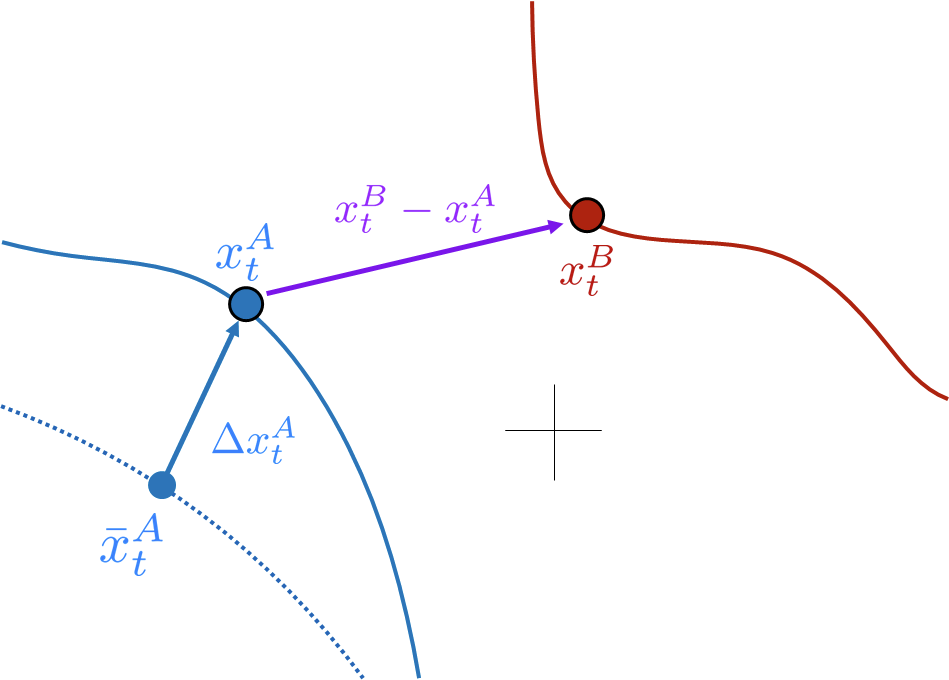}
    \caption{The state space considered in this analysis is the distance between Car A and Car B (purple arrow) and the distance between the lead car's trajectory and a nominal trajectory at each time step $t$. In this example, we presume that Car A is leading, and so $\Delta x$ is the difference between Car A's actual and nominal trajectories at each time step.}
    \label{fig:state-space}
\end{figure}

%
Specifically, as we are interested in better understanding a driver's priorities within an interaction, we choose a state space which contains both their own desired behavior as well.
as their position relative to the other vehicle.
We consider the state space 
$z = (x_B - x_A, \Delta x_p) \in \mathbb{R}^4$
Here, $x_B - x_A = (x_{B1} - x_{A1},x_{B2} - x_{A2}) \in \mathbb{R}^2$ is the planar distance between Car A and Car B in the lateral and longitudinal directions, respectively.
$\Delta x_p = ( \Delta x_{p1},\Delta x_{p2})  \in \mathbb{R}^2$ is the planar distance between the lead car's lateral and longitudinal position and a nominal trajectory.
For instance, if Car A is the first car to cross through the intersection, which we describe as ``leading,'' then $\Delta x_A$ is the difference between the Car A's actual position and a nominal trajectory for Car A, as shown in Fig. \ref{fig:state-space}.
The nominal trajectory is derived from a unicycle model completing a left turn through an intersection with lane-keeping constraints, solved via successive convex optimization \cite{10919603, acikmese2017}.
With this structure including both their distance from nominal and distance from the other vehicle, we believe we can interpret the different weights as prioritizing either the drivers desire to avoid the other vehicle or to follow their own planned trajectory.

We aim to learn the gains associated with two control inputs captured in the data: acceleration $a$ and angular velocity $\omega$.
In order to learn the gain matrix associated with these control inputs, we must accurately approximate the actual values of the control policies.
This requires fitting on $a$ and $\omega$ separately and assuming they are decoupled.
This gives the combined gain matrix for both control inputs the following structure:
\begin{equation}
\label{eqn:gain-matrix}
    G^T = 
    \begin{bmatrix}
        g^-_{a1} & g^-_{a2} & g^\Delta_{a1} & 
        g^\Delta_{a2} \\
        g^-_{\omega 1} & g^-_{\omega 2} & g^\Delta_{\omega 1} & 
        g^\Delta_{\omega 2}
    \end{bmatrix}
\end{equation}
where $
(g^-_{a1},g^-_{a2}) \in \mathbb{R}^2$ and $(g^-_{\omega 1},g^-_{\omega 2}) \in \mathbb{R}^2$ correspond to the gains on the $x_A - x_B$ components of the state space for each control, and $(g^\Delta _{a1},g^\Delta_{a2}) \in \mathbb{R}^2$ and $(g^\Delta_{\omega 1},g^\Delta_{\omega 2})  \in \mathbb{R}^2$ correspond to the gains on the $\Delta x_p$ components of the state space for each control.
Each row of $G^T$ is referred to as $G_a$ and $G_\omega$, respectively.

\begin{table}[t]
    \centering
   \caption{Number of Trajectories and Data Points in each distribution, categorized by Location, Lead Car, and Approach (App.), Intersection (Int.), and Exit Regions}
    \begin{minipage}{0.45\columnwidth}
        \centering
    \begin{tabular}[t]{p{0.5\columnwidth}p{0.3\columnwidth}}
    \toprule
     \textbf{Distribution}  & \textbf{\# of Data Points}\\ 
        \midrule
     ISR A App. & 1,413\\
     ISR A Int. & 1,264\\
     ISR A Exit & 598\\
     ISR B App.  & 646\\
     ISR B Int.  & 1,633\\
     ISR B Exit  & 834\\
        \bottomrule
        \end{tabular}
    \end{minipage}%
    \hfill
    \begin{minipage}{0.45\columnwidth}
        \centering
         \begin{tabular}[t]{p{0.5\columnwidth}p{0.3\columnwidth}}
         \toprule
     \textbf{Distribution}  & \textbf{\# of Data Points}\\
        \midrule
     NYC A App.  & 831\\
     NYC A Int.  & 1,097\\
     NYC A Exit  & 551\\
     NYC B App.  & 723\\
     NYC B Int.  & 2,298\\
     NYC B Exit  & 1,218\\
        \bottomrule
        \end{tabular}
    \end{minipage}

    \label{tab:distributions}
\end{table}
\subsection{Analysis of Gains}

We employ several methods for interpreting the gain matrices intuitively based on the rows separately and a singular value decomposition. 
In this preliminary work, we consider only linear feedback controllers, precluding us from learning nonlinear control structures that may be more realistic.  For example, intuitively we would expect the control action based on the relative position of the other vehicle to decrease with increased distance between the vehicles, i.e. have a nonlinear inverse distance relationship.  In cases such as this where the actual feedback law is likely to be nonlinear we expect the linear gains to capture some correct aspects, such as the correct sign of the control gain, but to be inaccurate in other aspects such as magnitude.

\subsubsection{Individual Row Interpretations}
Interpreting the gains that generate acceleration versus angular velocity requires slightly different intuition.
For both sets of gains, if $|g_{aj}|$ (or $|g_{\omega j}|$) are large then the control responds significantly to changes in $z_j$; if they are small then the control response is minimal. A positive gain $g_{aj} > 0$ means that acceleration increases as $z_j$ increases; a negative $g_{aj}<0$ means that acceleration decreases as $z_j$ increases.  A positive gain $g_{\omega j} > 0$ means that as $z_j$ increases, the angular velocity increases, i.e. the vehicle turns clockwise to the right; a negative gain $g_{\omega j} < 0$ means that as $z_j$ increases, the angular velocity decreases; i.e. the vehicle turns counterclockwise to the left.

\subsubsection{Singular Value Decomposition Interpretation}
\begin{figure}[t]
    \centering
    \includegraphics[width=0.75\columnwidth]{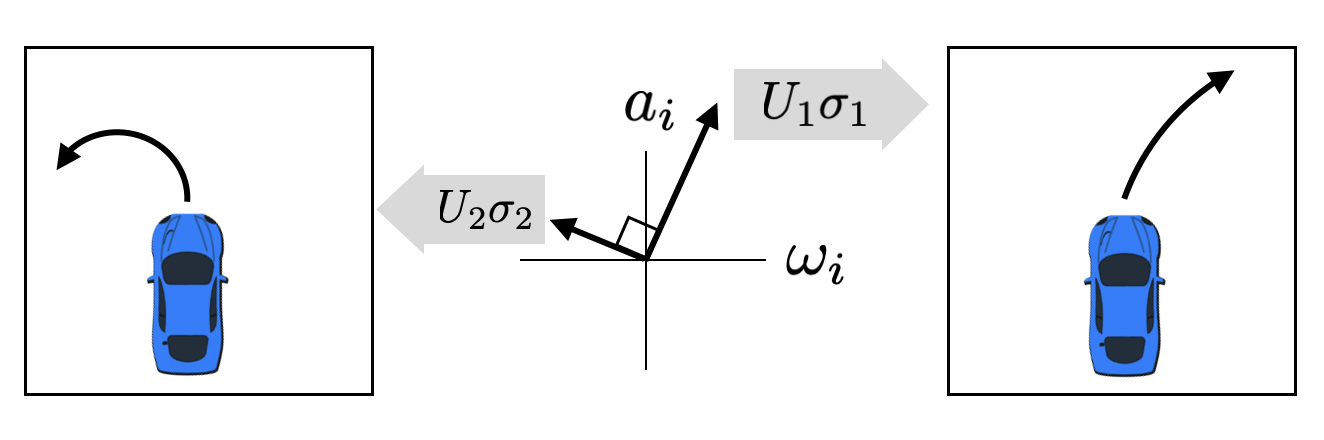}
    \caption{
    The SVD of the gain matrix $G$ can be intuitively interpreted as the most important components of the inputs acceleration $a$ and angular velocity $\omega$, with in turn have physical meaning associated with turning behaviors.}
    \label{fig:quivs}
\end{figure}

We can also investigate the gain matrix for the learned inputs using singular value decomposition.
This allows us to write $G^T$ in the form $G^T =  U\Sigma V^*$, where $U$ is a matrix with the output rotation, $\Sigma$ is a scaling factor, and $V^*$ is a rotation in the input space. For $G^T \in \mathbb{R}^{2 \times 4}$, we can specifically write 
\begin{align}
G^T 
& =
\begin{bmatrix}
| & | \\
U_1 & U_2  \\
| & | 
\end{bmatrix}
\begin{bmatrix}
\sigma_1 & 0 \\
0 & \sigma_2 
\end{bmatrix}
\begin{bmatrix}
- \ V_1^\top \ - \\
- \ V_2^\top \ - 
\end{bmatrix} \\
& =
\begin{bmatrix}
U_{a1} & U_{a2} \\
U_{\omega1} & U_{\omega 2} 
\end{bmatrix}
\begin{bmatrix}
\sigma_1 & 0 \\
0 & \sigma_2 
\end{bmatrix}
\begin{bmatrix}
V_{11} & V_{12} & V_{13} & V_{14}\\
V_{21} & V_{22} & V_{23} & V_{24}
\end{bmatrix} 
\end{align}
The columns $U_1,U_2$ are orthogonal directions in the control space that generate specific maneuvers with different combinations of linear acceleration, $U_{aj}$, and angular velocity, $U_{\omega j}$.  The vectors $U_j$ have unit length but if we scale them by $\sigma_j$, we can get a sense for how large that maneuver will be. Precisely, for $z = V_j$ (note here $z$ has unit norm), then the resulting control is $\Delta u = G^Tz = U_j \sigma_j$.  The elements of each row of $V^\top$ give the gains that generate a motion in the $U_j$ direction.
A cartoon representation of this analysis is shown in Fig. \ref{fig:quivs}.

\section{Results and Discussion}
\label{results}

\begin{figure}[t]
    \centering
    \includegraphics[width=0.4\linewidth]{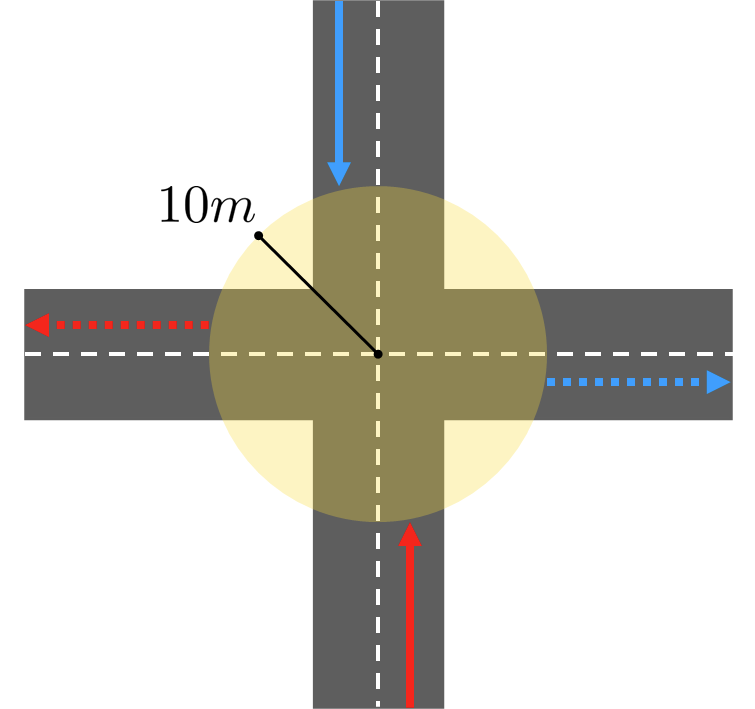}
    \caption{The state space is broken into three regions: the approach region denoted by a solid line for both Car A (red) and Car B (blue), the intersection denoted by the yellow circle with radius 10m, and the exit region denoted by a dotted line for Car A (red) and Car B (blue).
    }
    \label{fig:regions}
\end{figure}

\begin{figure}[t]
    \centering
    \includegraphics[width=0.8\linewidth]{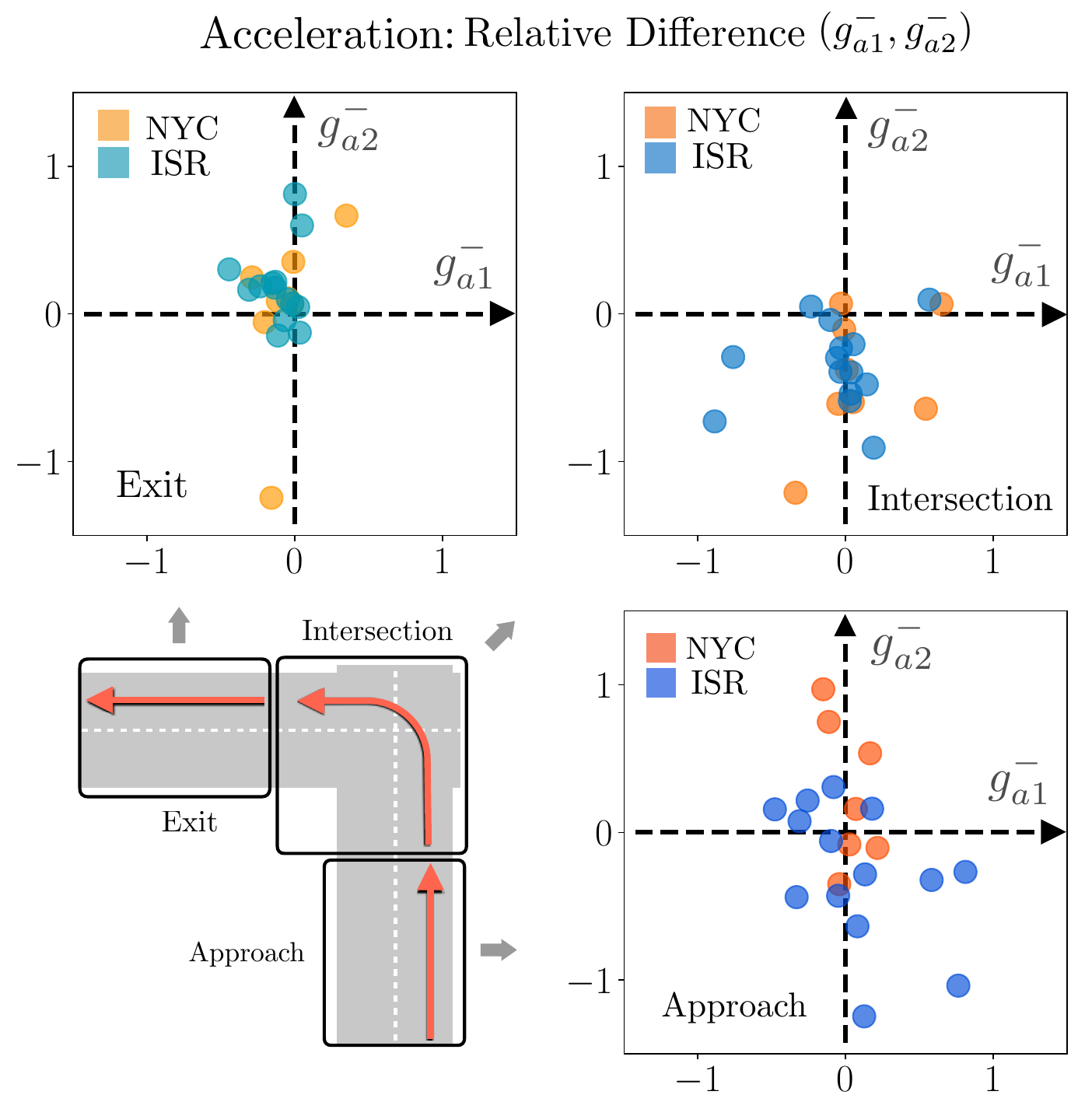}
    \caption{
    Relative position acceleration gains, $g_{a}^- \in \mathbb{R}^2$: Approach) Forward-backward relative distance causes deceleration in ISR population and acceleration in NYC population.  The ISR population also exhibits a wider spread of gains. 
    Intersection) Forward-backward relative distance causes deceleration for both populations, i.e. cars decelerate less when they are in closer proximity in the intersection. 
    Exit) Forward-backward relative distance causes acceleration for both populations.}
    \label{fig:acceldiff}
\end{figure}

\begin{figure}[t]
    \centering
\includegraphics[width=0.8\linewidth]{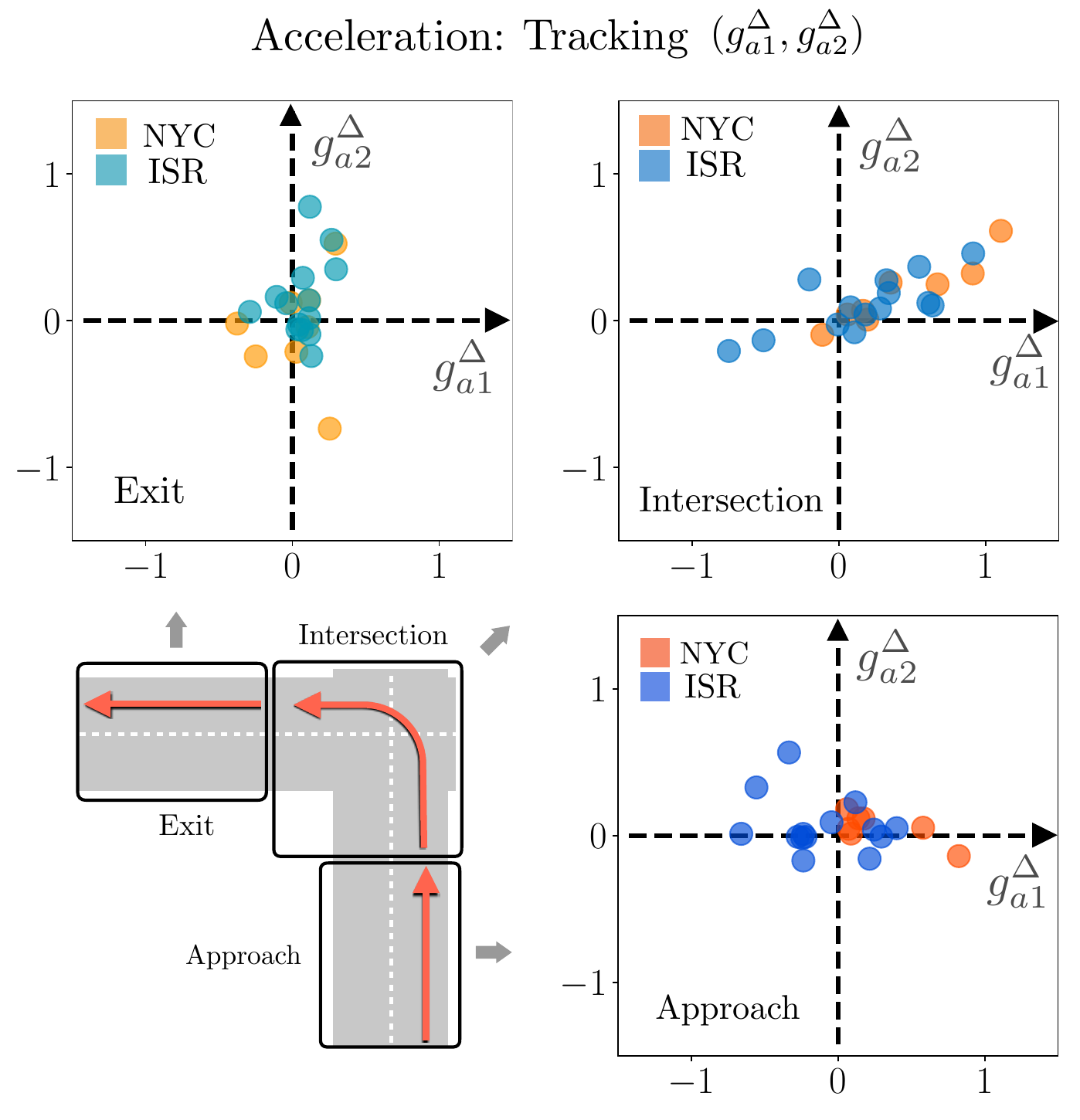}
    \caption{
    Tracking acceleration gains, $g_{a}^\Delta \in \mathbb{R}$: Approach \& Exit) Acceleration is sensitive to lateral deviations from the nominal trajectory.
    Intersection) Vehicles accelerate more if they are on the right side of their nominal trajectory (and decelerate if they are on the left side) consistent with making a left turn. 
    }
    \label{fig:acceltrack}
\end{figure}

\begin{figure}[t]
    \centering
\includegraphics[width=0.8\linewidth]{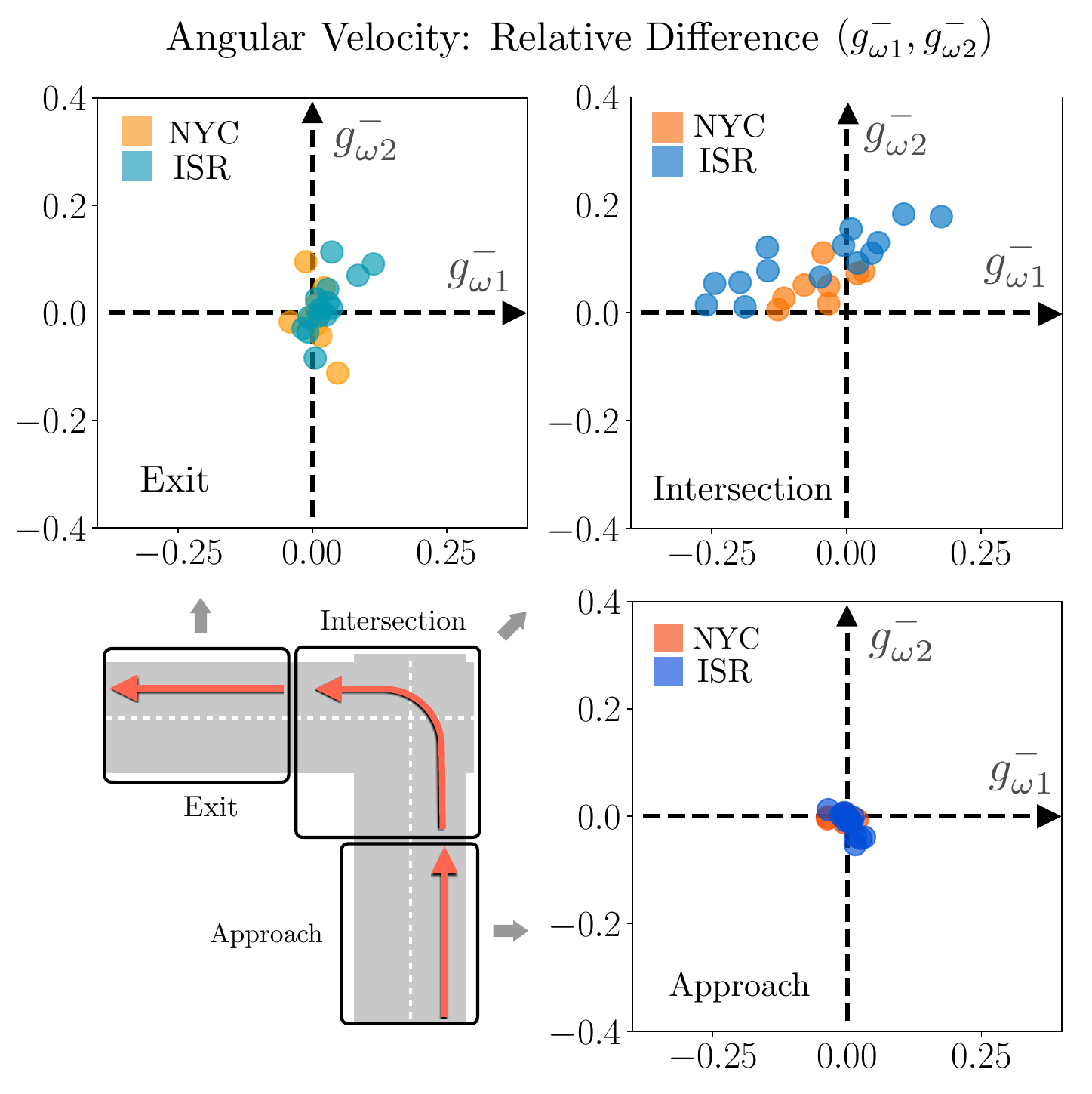}    
    \caption{
    Relative position angular velocity gains, $g_{\omega}^- \in \mathbb{R}^2$: Approach) Adjustments to orientation do not depend much on the other vehicle's position.  Intersection) Gains in the lower-right orthant indicate that agents turn left more readily when they are laterally far apart from the other agent and that they turn right more readily when the forward-backward relative distance is larger. }
    \label{fig:steerdiff}
\end{figure}

\begin{figure}[t]
    \centering
    \includegraphics[width=0.8\linewidth]{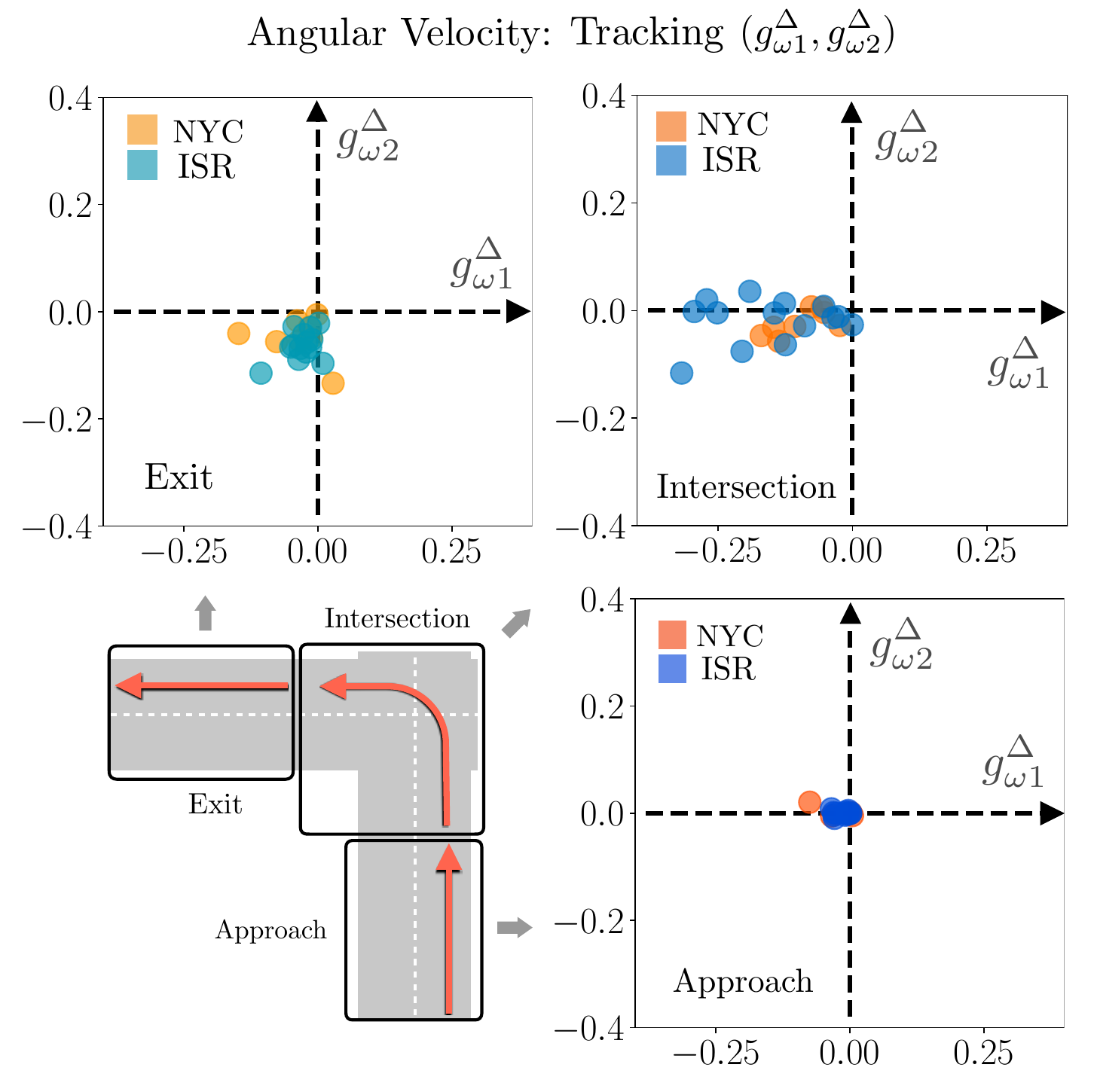}
    \caption{
    Tracking angular velocity gains, $g_{\omega }^- \in \mathbb{R}^2$: 
    Lateral deviation from nominal to the right causes left-turn to correct course. (Similarly, a lateral deviation to the left causes a right-turn.)  This effect is evident in all three regions but least pronounced in the approach region (with very small gains) and most pronounced in the intersection.  This correction behavior also varies most widely between individuals in the intersection and more widely among ISR drivers than NYC drivers.}
    \label{fig:steertrack}
\end{figure}

\begin{figure}[t]
    \centering
    \includegraphics[width=0.99\columnwidth]{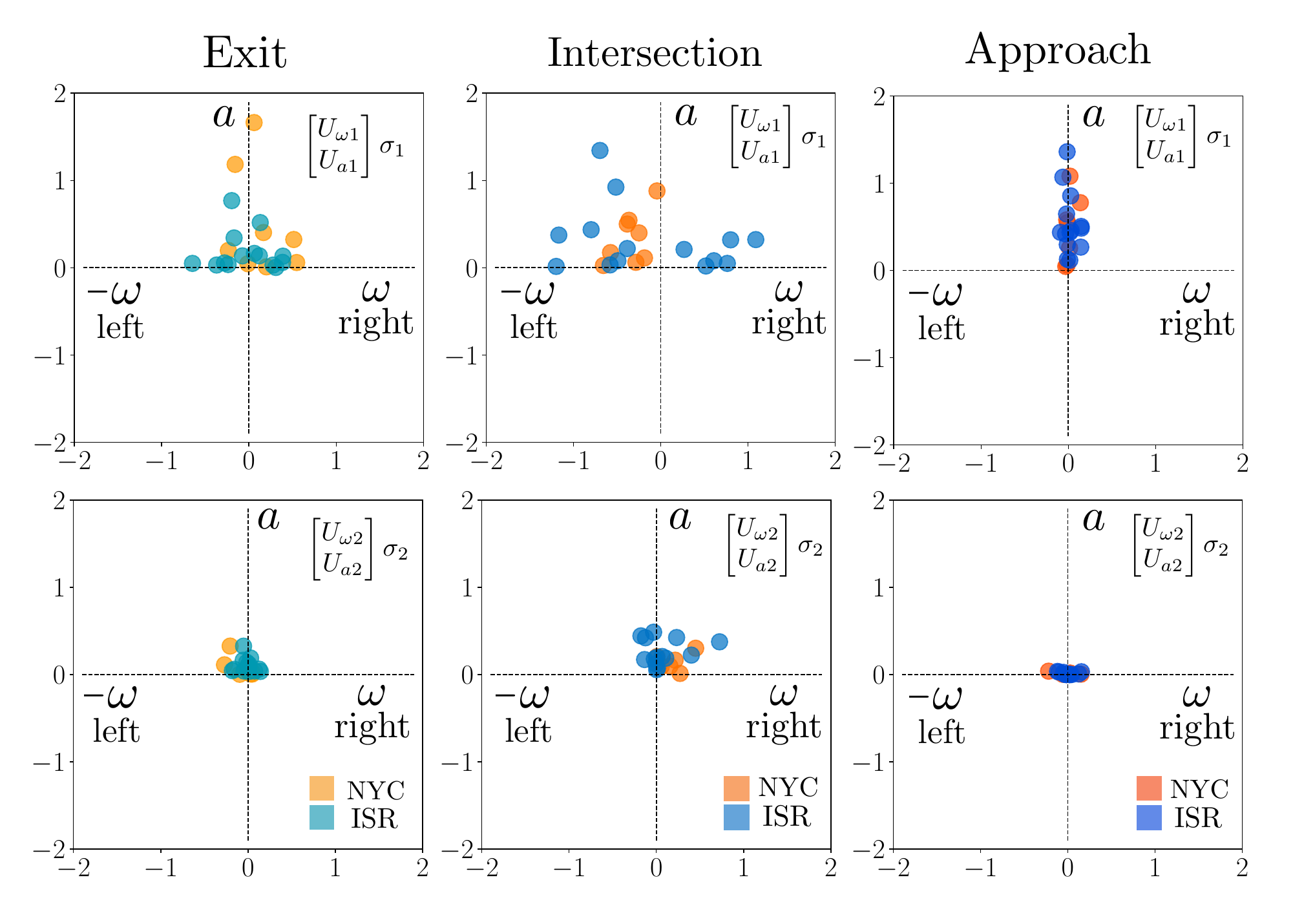}
    \caption{
    The columns of $U$ of the SVD of the gain matrix $G$ can be intuitively interpreted as the most important components of the inputs acceleration $a$ and angular velocity $\omega$, with in turn have physical meaning associated with turning behaviors.}
    \label{fig:svd}
\end{figure}

\subsection{Data selection and pre-processing}


As discussed in Section \ref{experiment}, the data used to fit the feedback laws consist of joint trajectories of drivers navigating an unsigned intersection.
We only consider data points in which an interaction has taken place, which we determine as when both cars are within an approach region of 25 meters of the intersection.
We consider this approach region because 1) it is when both cars are within sight of the intersection, and 2) we found it to have high accuracy in identifying interactions as compared to researcher-identified interactions, as discussed in \cite{10919603}.
In addition, we consider one specific type of navigation through the intersection where both drivers are instructed to turn left.
Any trials in which either driver does not turn left through the intersection due to variability of human behavior within the simulator are excluded from the analysis, as the learned controller would not match what was expected in the scenario.
This stipulation leads to 21 interactions being excluded from the analysis, with 42 remaining.

The joint trajectory starts when both vehicles are within the 25 meter radius circle defining the approach region of the intersection.
We hypothesize that for a linear control formulation to be effective, different regions of the state space will have different linear gains; that is, the linear controller is piecewise-affine over the entire state space.
We consider the following portions of the state space individually when fitting the feedback law, which we call the approach region, intersection, and exit region.
The approach region is defined as when the vehicle whose policy we are fitting is between 10 to 25 meters from the center of the intersection.
The intersection is defined by a circle with a radius of 10 meters centered at the origin.
We consider the intersection region to capture both the behavior immediately reaching the intersection and through the intersection, as these are the most likely to be impacted by another vehicle. The radius of 10 meters was chosen heuristically as it appeared to capture the most of the immediate approach behavior and turning behavior; however, future work will consider more rigorous methods for identifying the different regions of the driving space.
The exit region is the region where the lead car has passed through the intersection entirely; that is, the car has left the circle of radius 10 meters.
Fig. \ref{fig:regions} depicts how the trajectory is divided for analysis for both cars A and B.

The trajectories are divided into eight different distributions: by which car arrives at the intersection first, Car A Leads or Car B Leads (denoted as ``A" or ``B", respectively), and then further by the location the data was collected in (``ISR" or ``NYC").
Overwhelmingly, it was found that the car that arrived at the intersection first took right-of-way, leading to drastically different joint behavior depending on when the cars arrived.
Because we wish to identify the feedback law used when completing a left turn in an interaction, we consider which car arriving at the intersection first eliciting different behaviors from the drivers.
Table \ref{tab:distributions} contains the total number of data points, corresponding to time steps, from the joint trajectories used for each distribution.
We learn the gain matrix for each distribution of data separately.

We also learn gain matrix for each individual driver.
This allows us to better understand the heterogeneity within and between the populations of drivers.
Of the trajectories with interactions, 14 were from Israel with Car A Leading, 13 were from Israel with Car B Leading, 8 were from New York with Car A Leading, and 17 were from New York with Car B Leading.
These trajectories do not necessarily have the same length, so the gains for the individual trajectories may be different than those learned when pooling all the data points from the trajectories.

Lastly, we normalize the input data used in the prediction to be between -1 and 1 to facilitate accuracy of the GP regression.
80\% of the data are used for training, and the remaining 20\% are used for validation and estimating the generalization error.
The mean squared error for each of the distributions of data is shown in Table~\ref{tab:mse}.
The nonlinear inclusion factor, $\beta$, of each distribution of data points is also included in Table~\ref{tab:mse}.
We performed K-fold cross validation \cite{hastie01statisticallearning} using 5 folds and did not see significant differences in the MSE metric, with a minimum, average, and maximum of 2.542e-17, 0.107, 4.287, respectively, suggesting that the model is robust.

As the hyperparameter weighting on the squared exponential kernel was tuned during fitting, there were instances where the nonlinear kernel outweighed the linear one.
We observed significant nonlinearities in the ISR B and NYC B intersection and NYC B exit regions  which appeared to be due to ``stuttering'' in the acceleration as drivers showed hesitation in progressing through and leaving the intersection.  Such behavior is clearly beyond what a linear model can capture.  Future work will involve more in depth analysis of the nonlinearities and their implications for AV design.

In addition, we compared our proposed model to ones that only use linear or squared-exponential priors instead of both. The results are summarized in Table~\ref{tab:model_compare}.
While the linear prior model and the squared-exponential model appear to do better based on the MSE metrics, the linear model does not capture known nonlinearities in behavior and the squared exponential model is prone to overfitting.
By combining these two priors, we believe that we achieve an interpretable and robust structure for our controller without sacrificing much in terms of accuracy.

\begin{table}[t]
    \centering
        \caption{MSE Comparison of Linear, Squared-Exponential, and Combined Models}
    \begin{tabular}{c|ccc}
    \toprule
       \textbf{MSE} & \textbf{Linear} & \textbf{Squared-Exp} & \textbf{Combined}  \\
    \midrule          
    \textbf{Min} & 1e-4 & 1e-4 & 0.032e-4 \\
    \textbf{Avg} & 0.027 & 0.002 & 0.110 \\
    \textbf{Max} & 0.099 & 0.006 & 0.444 \\
    \bottomrule
    \end{tabular}
    \label{tab:model_compare}
\end{table}

\begin{table*}[t]
    \centering
    \caption{Mean-Squared Error and $\beta$ for each distribution by Location, Lead Car, Region, and Control Input}
        \begin{tabular}{p{0.49\columnwidth} p{0.2\columnwidth} p{0.17\columnwidth}}
        \toprule
        \textbf{Distribution} & \textbf{MSE} & $\beta$ \\ 
        \midrule
        ISR A Approach Accel. & 0.142 & 0.075\\
        ISR A Approach Ang. Vel. & 3.20e-05 & 1.0e-2\\
        ISR A Intersection Accel. & 0.228 & 0.588\\
        ISR A Intersection Ang. Vel. & 0.006 & 1.0e-2\\
        ISR A Exit Accel. & 0.430 & 0.247\\
        ISR A Exit Ang. Vel. & 0.002 & 1.0e-2\\
        ISR B Approach Accel. & 0.108 & 0.088\\
        ISR B Approach Ang. Vel. & 1.2489e-04 & 1.0e-2\\
        ISR B Intersection Accel. & 0.444 & 1.147\\
        ISR B Intersection Ang. Vel. & 0.004 & 3.020e-2\\
        ISR B Exit Accel. & 0.154 & 0.383\\
        ISR B Exit Ang. Vel. & 0.002 & 1.0e-2\\
        \bottomrule
        \end{tabular}
    \hfill
    \begin{minipage}{0.9\columnwidth}
        \centering
        \begin{tabular}{p{0.5\columnwidth} p{0.2\columnwidth} p{0.14\columnwidth}}
        \toprule
        \textbf{Distribution} & \textbf{MSE} & $\beta$ \\ 
        \midrule

        NYC A Approach Accel. & 0.126 & 0.094\\
        NYC A Approach Ang. Vel. & 5.140e-05 & 1.0e-2\\
        NYC A Intersection Accel. & 0.165 & 0.255\\
        NYC A Intersection Ang. Vel. & 0.002 & 1.0e-2\\
        NYC A Exit Accel. & 0.190 & 0.320 \\
        NYC A Exit Ang. Vel. & 1.0e-3 & 1.0 e-2 \\
        NYC B Approach Accel. & 0.109 & 0.067\\
         NYC B Approach Ang. Vel. & 7.53e-05 & 1.0e-2\\
         NYC B Intersection Accel. & 0.336 & 3.134\\
         NYC B Intersection Ang. Vel. & 0.002 & 1.0e-2\\
         NYC B Exit Accel. & 0.185 & 9.001 \\
         NYC B Exit Ang. Vel. & 1.0e-2 & 1.0e-2\\
        \bottomrule
        \end{tabular}
    \end{minipage}

    \label{tab:mse}
\end{table*}

\subsection{Interpreting Driving Behaviors from Gain Matrices}

Examining the gains learned for each individual driver reveals intuitive driving behaviors for both the acceleration and angular velocity
even with the given relatively simple state space.  These results 
provide a promising indication that human behavior can be modeled as a linear feedback law which can then be leveraged to predict human behavior. 

In Fig. \ref{fig:acceldiff} within the intersection region, we can identify that drivers slow down as they approach another vehicle in front of them.
We see this via the negative sign on $g^-_{a2}$, which indicates that the acceleration decreases when the distance between the vehicles increases.
This matches the expected behavior that drivers accelerate less and even decelerate as they reach the intersection.


Similarly, in Fig. \ref{fig:steertrack}, we can see that drivers typically do not turn on straight stretches of road, such as the approach and exit regions in our scenario.
In the approach region in Fig. \ref{fig:steertrack}, both the longitudinal and lateral gains are small when considering how closely a vehicle is tracking its desired trajectory; essentially, if a vehicle is going straight on a straight section of road, the driver does not prioritize turning.
As the vehicle is primarily driving straight in the approach region, we would expect little variation in the angular velocity, so long as the vehicle is lane-keeping, as is reflected by the small magnitude of the gains.



We can also discern a clear tendency for a driver to track their nominal trajectory.  In Fig. \ref{fig:steertrack} in the intersection region, the gain that reacts to lateral deviation from nominal, $g_{\omega 1}^\Delta$, is negative.  
This
suggests that the vehicle is turning to the left (negative angular velocity) to course-correct on to their expected path when the car is to the right of its expected trajectory. ($\Delta x_{p1} >0$).
Similarly deviating to the left of nominal ($\Delta x_{p1} <0$)
will cause a course correction to the right.
Also in Fig. \ref{fig:steertrack} in the exit region, the gain $g_{\omega 2}^\Delta$ is negative which corresponds to similar tracking behavior based on lateral deviation from nominal since now the car is driving to the left as opposed to straightforward, and so the control actions are negative in the y-axis rather than the x-axis as in the intersection region.


We can also identify differences in turning maneuvers by drivers based on their gain matrices.
Specifically, we identify that some drivers speed up while turning, while others slow down.
In Fig. \ref{fig:acceltrack} in the intersection when the drivers are turning left, the gain in the positive orthant, indicating that drivers speed up when the state is positive, namely, they are to the right and in front of their nominal trajectory.
We interpret this as the driver speeding up to get back-on-track when they are taking their turn too wide to the right.
Conversely, the gains in the negative orthant suggest that drivers slow down when they are behind or to the left of their nominal trajectory, i.e. cutting their left turn too sharply.

\subsection{Implications for Socially Responsive AV Control Design}

We observe differences in gains between groups of drivers at the two locations in which the studies were performed.
First, we consider the acceleration in the approach region.
In Fig. \ref{fig:acceldiff}, the gains in the approach region indicate that longitudinal relative distance between vehicles causes deceleration in ISR population and acceleration in NYC population. 
We believe that this suggests NYC drivers speed up in the approach region to take advantage of the empty space in front of them.
Conversely, ISR drivers may be displaying hesitation while approaching the intersection, as described in \cite{10919603}.
We validated that the difference in acceleration gains $g_{a2}^-$ is due to differences between populations, by performing a permutation test with 1,000 permutations, that compared the difference between the means of each group, with  
a p-value of 0.02.

Second, we consider the steering angle while drivers perform left turns.
In Fig. \ref{fig:steertrack}, in the intersection, the ISR population has more variability in the gains they use to correct steering; many of them also have larger gains than the NYC drivers, indicating they may be more aggressive in correcting toward the nominal path, and prioritizing a smoother turn shape.
We validated this result through a permutation test, that compared differences in the variance of the distributions, and found p-values near zero.  
For socially responsive AV design, controllers that reflect these differences may be important: vehicles that 
accelerate towards an intersection would be less expected and interpretable by ISR drivers; similarly, vehicles that cannot hew to a narrow range of left turns may be well-received in NYC, but acceptable in ISR.

\subsection{Characterizing Controllers for Maneuvers via SVD}
Lastly, we note that we can employ SVD analysis on the gain matrices for each population to characterize maneuvers, by identifying patterns in control gains associated with certain regions of the state-space.  As an example,
Fig.\ref{fig:svd} shows control patterns associated with the left turns during different portions of the intersection.

The left singular vectors $U$ could be used to construct low-dimensional representations of human driving behaviors to facilitate the prediction of human behavior for AV control design. 
The right singular vectors $V^*$ (not shown) could be used to determine features in the state space that drivers are most sensitive to when making control decisions for estimation and control design. These linear features could serve as a basis for dimensionality reduction techniques for designing low dimensional linear feedback controllers in larger state spaces. 

\section{Conclusion}
\label{conclusion}

In this work, we present a model of interpreting human driver interaction control policies.
We fit this model to naturalistic interaction data using GP regression and find that we can identify patterns in how drivers prioritize their position relative to other cars versus their own desired trajectory.
We apply singular value decomposition to identify different types of maneuvers performed by drivers from the structure of the learned linear gain matrix.
In future work, we can use this method to explore other state space configurations and analyze the important basis functions using principal component analysis.
These types of insights are instrumental in developing socially-aware autonomous vehicles.

\bibliographystyle{IEEEtran}
\bibliography{bibliography,rkhs,humandecisionmaking}

\end{document}